\documentclass{ws-ijmpa}

\begin{document}
\hyphenation{mo-dels}

\markboth{Nicolas Bock, Thomas J. Humanic}
{Quantitative Calculations for mini-Black Hole production at the Large Hadron Collider}

%
\catchline{}{}{}{}{}
%

\title{QUANTITATIVE CALCULATIONS FOR BLACK HOLE PRODUCTION AT THE LARGE HADRON COLLIDER} 

\author{NICOLAS BOCK\footnote{bock@mps.ohio-state.edu}  \hspace{1mm} and THOMAS J. HUMANIC\footnote{humanic@mps.ohio-state.edu}}

\address{Department of Physics, The Ohio State University\\
 191 W. Woodruff Avenue, Columbus, Ohio, 43210\\
United States of America
}

\maketitle

\begin{history}
\received{\today}
\end{history}

\begin{abstract}

The framework of Large Extra Dimensions provides a way to explain why gravity is weaker
compared to the other forces in nature. A consequence of this model is the possible production of 
D-dimensional Black Holes in high energy p-p collisions at the Large Hadron Collider. The present work uses the CATFISH Black Hole 
generator to study quantitatively how these events could be observed in the hadronic channel at mid-rapidity using a particle tracking detector. 


\end{abstract}

\ccode{PACS numbers: 11.25.Hf, 123.1K}

\section{Introduction}

With the Large Hadron Collider (LHC) opening in late 2008, physics at the 14 TeV center of mass 
scale will be probed in p-p collisions. Although searching for the Higgs particle will be one of the main
research focuses, other interesting physics, such as the possible formation of D-dimensional Black Holes (BH) in these
collisions will be another direction of intense research.
This phenomenon is predicted by the framework of
Large Extra Dimensions (LED) in the Arkani-Hamed, Dimopolus and Dvali (ADD) model \cite{ADD1,ADD2}. This model is a proposed solution to the Hierarchy Problem, the question of why gravity is much weaker than the other forces in nature. 
The scale for gravity is characterized by the Planck mass, $M_P\approx 10^{19}$ GeV, whereas the strong and the weak forces have a scale on the order of 1 GeV/fm and 100 GeV, respectively. 

The approach to solve the problem in the ADD model uses the following four assumptions: 1) the hierarchy of the forces in nature only exists in 3+1 dimensions, but not in higher dimensional space-time with $D$=3+1+$n$ dimensions, 2) only gravity is allowed to propagate in the $n$ extra dimensions through gravitons, 3) the LED are ``compact" or finite, but too small to be detected until (possibly) now, 4) Standard Model particles only propagate in 3+1 dimensional space-time, which is embedded in $3+1+n$ higher dimensional space-time. Gravity seems to be weaker in 3+1 dimensions because it is diluted in 3+1+$n$ higher dimensional space. 
The Planck mass in a theory with $n$ LED is given by

\begin{equation}
M_P \approx \left(\frac{\hbar}{2 \pi r_c}\right)^{\frac{n}{n+2}} M_{P0}^{\frac{2}{n+2}},
\label{PlanckMass}
\end{equation}

\noindent where  $M_{P0} = \sqrt{\hbar/G}=1.22 \times 10 ^{19}$ GeV is the Planck mass in $D$=3+1 dimensions
and $r_c$ is the compactification length of the extra dimensions. The compactification length is a free parameter of the theory, and it can be seen 
\cite{HumanicAlice} that for a given Planck mass,  it decreases as the number of extra dimensions increases, for example for $M_P$ = 1 TeV $r_c$ is close to 1 fm for $n$=7. For fixed $r_c$ the Planck scale is lowered as the number of LED increases and the hierarchy of the forces is largely suppressed \cite{HumanicAlice}.

One of the consequences of this model is the formation of $D$-dimensional BHs smaller than the size of the extra-dimensions and centered on the brane. The gravitational radius of the $D$-dimensional BHs is up to  $10^{32}$ times larger than that of a usual BH in 3+1 dimensions with the same mass. This considerably increases the possibility of creating such an object at the LHC.

The main purpose of this work is use the CATFISH \cite{catfish} Monte Carlo (MC) code to study the possible signatures of BHs formed in p-p collisions, which could be observed in the hadronic channel at mid-rapidity using a charged particle tracking detector.  CATFISH is a MC code based on the ADD model and is used to generate BH events at the LHC with center of mass energy of 14 TeV.
It is similar to other MC generators like TRUENOIR \cite{truenoir} and CHARYBDIS \cite{charybdis}, but it is more up to date 
since it uses the most recent theoretical results of BH formation and evolution. The  updates include inelasticity effects during the BH 
formation phase, exact field emissivities, corrections to Hawking semi classical evaporation phase, BH recoil on 
the brane, and additional final BH decay modes (including remnants). CATFISH links to the PYTHIA \cite{pythia} MC code 
to simulate the evolution of the decay products of the BH into Standard Model particles. The flexibility of 
CATFISH allows the study of the different theoretical models of BH formation. 
A comparison between the features of CATFISH and CHARYBDIS has been carried out before \cite{Gingrich}
and a similar study of the observables of BH at the LHC has also been done using the CHARYBDIS Code \cite{HumanicAlice,Humanic}.

CATFISH includes the following three models to calculate the BH evolution: 
\begin{itemize}
                \item  \textbf{No Gravitational Loss model (NGL):} This model works in the semi classical limit of BH formation. The hoop conjecture \cite{Hoop} is used to estimate the possibility of BH formation in a particle collision. It states that an apparent horizon is formed in $D$=4 dimensions if a mass $M$ is compacted in a region with a circumference $C$ such that 
                \begin{equation}
                    H_D\equiv \frac{C}{2\pi r_h(M)}\leq 1,
                    \label{hoop conjecture}
                \end{equation}
                where $r_h(M)$ is the Schwarzschild radius for the mass $M$ given by
                                \begin{equation}
                                                r_h(M)=\left(\frac{16\pi G_D M}{(D-2)\Omega_{D-2}} \right)^{1/(D-3)}.
                                                \label{Schwarzschild Radius}
                                \end{equation}
      Here $\Omega_{D-2}$ is the volume of the ($D-2$)-sphere and $G_D$ is the gravitational constant.
                 The impact parameter $b$ of two colliding partons has to be smaller than $r_h(M)$ to produce a BH.  
                                The cross section is given approximately by the geometrical Black Disk (BD) cross section
                                \begin{equation}
                                                \sigma_{BD}=\pi R^2(s,n)\Theta[R(s,n)-b]
                                                \label{Black Disk}
                                \end{equation}  
                         where $R$ is the horizon radius and depends on the center of mass energy $\sqrt{s}$,  and the number of extra 
                         dimensions $n$. The BH mass is equal to the center of mass energy of the partons forming the BH. This is the same model used in TRUENOIR and CHARYBDIS.

                \item \textbf{Yoshino-Nambu model (YN)} \cite{YN,Catfish16}: 
                This model uses the Trapped Surface approach, which gives a bound on the inelasticity of a collision by modeling two                             
                incoming partons as Aichelburg-Sexl shock waves\cite{Aichel-Sexl}. The apparent horizon is found in the union of the two shock 
                waves. The condition for BH formation is better described in higher dimensionalities by the volume conjecture than by the hoop                  conjecture, and is given by 
                        \begin{equation}
                        \textit{H}_D\equiv \left[ \frac{V_{D-3}}{\Omega_{D-3}r_h^{D-3}(s,n)}\right]^{1/ (D-3) }\leq 1,
                        \label{volume conjecture}
                        \end{equation}
                
                where $\Omega_{D-3}$ is the volume of the ($D-3$)-sphere and $V_{D-3}$ is the characteristic ($D-3$)-dimensional volume
                of the system. The volume conjecture reduces to the hoop conjecture in $D$ = 4 dimensions.
                The cross section is calculated in this model as
                \begin{equation}
                                \sigma_{BH production}= F(D)\pi r_h^2(s,n),
                                \label{Sigma YN}
                \end{equation}

                where $F(D)$ is a numerical factor close to unity. The BH mass is less than the center of mass energy of the 
                partons forming the BH due to emission of gravitons and it depends on the impact parameter. 
                
                \item \textbf{Yoshino-Rychkov model (YR)} \cite {YR}: This model is an improved version of the YN model in that the apparent
                 horizon is constructed from a slice of the future light cone in the shock collision plane.  The slice is to the future of 
                 that used in the YN model. The condition for BH formation is also given by the volume conjecture and the cross section 
                 calculation is similar to the calculation for Equation \ref{Sigma YN}. The BH mass is also reduced in comparisson
                 to the BH mass in the NGL model.
\end{itemize}

The event simulation in CATFISH occurs in three steps:
\begin{romanlist}[(ii)]
\item The initial BH mass is sampled from the differential cross section.
\item The BH is decayed through the Hawking mechanism and final hard events.
\item The unstable quanta emitted are hadronized or decayed instantaneously by PYTHIA, except top quarks, which are decayed as
$t\rightarrow bW$ first. 
\end{romanlist}

The results of our calculations are presented in the following section. 
 
\section{Results and Calculations}

The signatures of BH production are studied in this section, first for completely decaying BHs and then for BH remnants.
The results of our calculations are shown in Figures 1 - 12. 
\subsection{BH signal using the hadronic channel}

The hadronic channel is considered as a possible method to detect BHs.
To be conservative, the first-year luminosity, the transverse momentum resolution and the rapidity are assumed to be $L = 10^{31}$ cm$^{-2} $s$^{-1}$,  0.1 GeV/c  $< p_T < $  300 GeV/c and -1 $< y <$ 1 respectively (e.g. these characteristics are similar to those of the central tracking detectors of the ALICE experiment \cite{ALICE} at CERN). 

The transverse momentum distribution $\frac{1}{p_T} \frac{dN}{dp_T}$ \cite{HumanicAlice,Humanic}  
for BH events flattens more than in the QCD background events at $p_T >$ 200 GeV/c  for $M_P$=1 TeV and at higher $p_T$ for larger values of $M_P$. This would make the detection of BH events difficult for tracking detectors since the momentum resolution of these detectors is poor for large $p_T$. To overcome this problem another observable can be used instead, namely the sum of the transverse momentum of all charged hadrons in each event, calculated using 

\begin{equation}
P_T=\sum_{i}p_{T_i},
\label{Summed PT}
\end{equation} 

\noindent  where $i$ runs over all charged hadrons in one event.
CATFISH is used to generate BH events and for each the $P_T$ is calculated and histogrammed in bins of size of 0.1 TeV/c. The total number of counts per bin for an arbitrary period of time $N$ can be calculated  using the following relation:

\begin {equation}
N = \frac{N_{bin} \sigma L t}{N_{Tot}} ,
\label{NumEvents}
\end {equation} 

\noindent where $N_{bin}$ is the number of counts in a particular bin, $N_{Tot}$ is the total number of events generated, and
$\sigma$ is the cross section. For the each simulation in this work, $N_{Tot}=10^5$ events were generated and a period of time of four months was chosen, so that $t \approx 10^7s$.

\begin{figure}[pb]
\centerline{\psfig{file=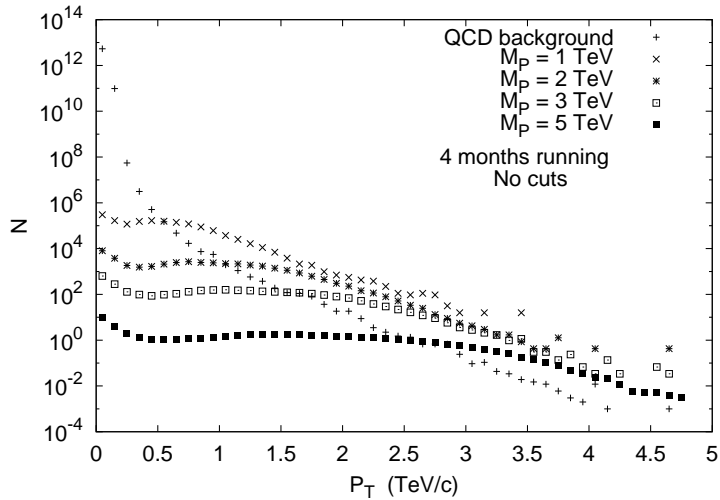,width=7.0cm,angle=270}}
\vspace*{8pt}
\caption{Comparison of the $P_T$ for different values of $M_P$ with the QCD background. 
        Model: NGL
\label{SPt in CH NoCuts}}
\end{figure}

In Figure \ref{SPt in CH NoCuts} $N$ vs. $P_T$  is plotted
for the NGL model with different values of the Planck mass, together with a normal 
QCD background run from PYTHIA. This run is a mixture of six different runs changing the hardness
of the $2 \to 2$ parton collision. It is seen that at low $P_T$ the normal QCD events dominate, but above 0.5 TeV the $M_P$ = 1 TeV 
curve takes over. This is a possible method  
for detecting the BHs: i.e, noticing the discontinuous change in the slope $\frac{dN}{dP_T}$ from normal QCD background events to BH events. 
For higher values of the Planck mass the curves intersect the QCD background at higher $P_T$ values. The BH production
becomes more suppressed but it can be seen that even at a $M_P$ of 5 TeV it would still be possible to 
detect a signal under these running conditions.
Furthermore, by obtaining experimentally the $P_T$ value at which the change in slope happens, the actual value of the Planck mass could be determined.

\begin{figure}[pb]
\centerline{\psfig{file=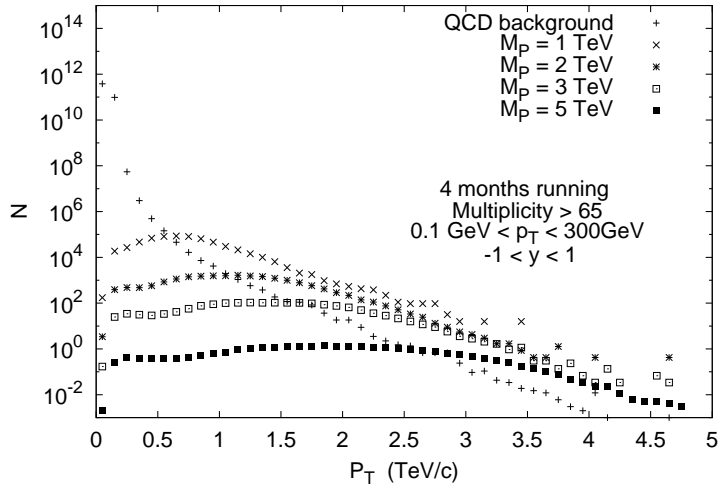,width=7.0cm,angle=270}}
\vspace*{8pt}
\caption{Comparison of the $P_T$ with applied cuts for different values of $M_P$ and the QCD background. Model: NGL. 
\label{SPt in CH WithCuts}}
\end{figure}

Figure \ref{SPt in CH WithCuts} plots the same quantities as in Figure \ref{SPt in CH NoCuts} except that various cuts are applied to simulate 
the acceptance of a realistic tracking detector. The cuts used are on rapidity ($-1< y < 1$), transverse momentum ( 0.1 GeV/c  $< p_T < $  300 GeV/c)
and multiplicity in the rapidity cut ($m>$65). As seen the BH signal is not degraded by these cuts. Its effect is particularly visible in lowest $P_T$ bins where the number of counts is reduced. These cuts will be applied in all subsequent plots.



\begin{figure}[pb]
\centerline{\psfig{file=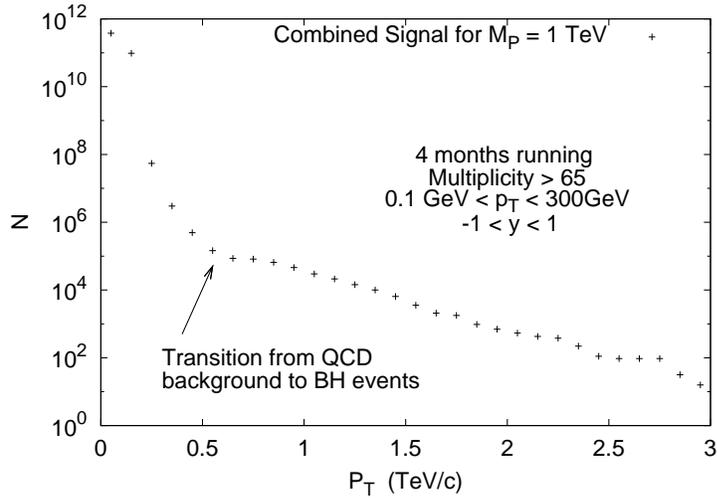,width=7.0cm,angle=270}}
\vspace*{8pt}
\caption{Combined signal of the QCD background and the BH signal for a $M_P$ = 1 TeV. Model: NGL. 
\label{BHsignal}}
\end{figure}

Figure \ref{BHsignal} plots the combined signal of the QCD background and the BH contribution, as one would obtain from the experiment. It can be seen that at $P_T\approx$ 0.5 TeV/c,  the slope of the graph changes abruptly indicating the transition from normal QCD events to BH events.

\begin{figure}[pb]
\centerline{\psfig{file=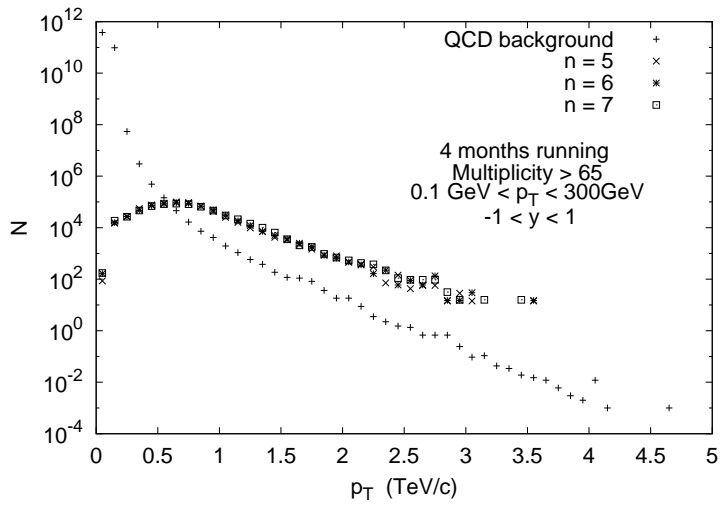,width=7.0cm,angle=270}}
\vspace*{8pt}
\caption{Comparison of the $P_T$ for $M_P$ of 1 TeV for different number of extra dimensions and the QCD background. 
\label{ExtraDim}}
\end{figure}

In Figure \ref{ExtraDim} the effect of varying the number of extra dimensions is studied. As can be seen 
there is no substantial difference in our results if five, six or seven extra dimensions are used in the simulations.
String Theory \cite{EdWitten} favors n=7 , so we have chosen to use this value for all of the results presented in the rest of the study.

\begin{figure}[pb]
\centerline{\psfig{file=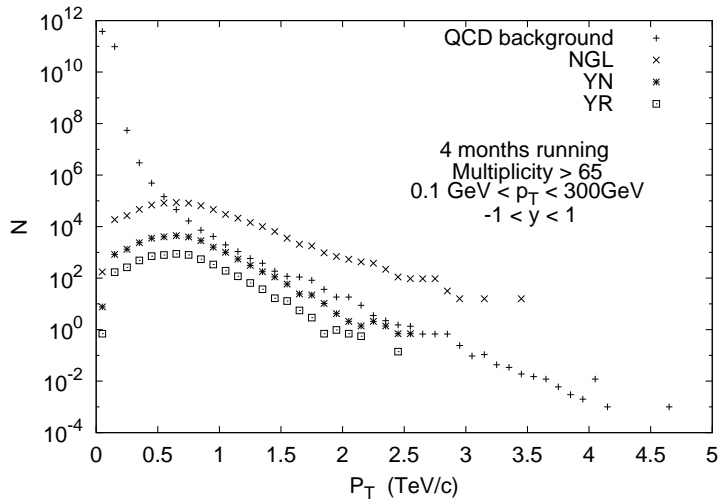,width=7.0cm,angle=270}}
\vspace*{8pt}
\caption{Comparison of the three different BH formation models for $M_P$ = 1 TeV: No Gravitational Loss (NGL), Yoshino-Nambu (YN) and Yoshino-Rychkov(YR).
\label{NGL-YN-YR}}
\end{figure}

In Figure \ref{NGL-YN-YR} the $P_T$ is plotted for the three theoretical models of BH formation: NGL, YN and YR. 
The NGL model has the largest cross section of the three models since it uses the semi classical approximation to BH formation. The YN and YR models take into account the gravitational energy loss at formation and this decreases the cross 
section.  It can be seen that only the NGL curve is above the QCD 
background, meaning that if BHs are formed according to the YN or YR models it would not be possible to detect any 
using the $P_T$ method.  From this figure we assert that the NGL 
model is an upper bound to the formation of BHs, and the YN and YR are the lower bounds for the known models. 

\begin{figure}[pb]
\centerline{\psfig{file=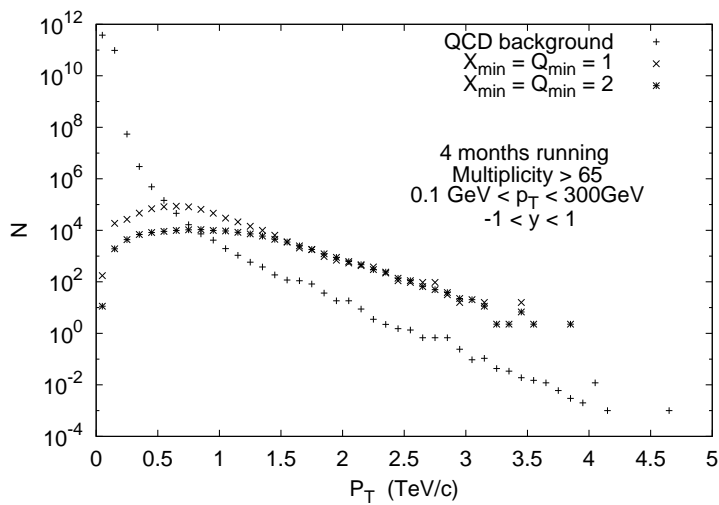,width=7.0cm,angle=270}}
\vspace*{8pt}
\caption{Comparison of two values of the BH mass at formation (Xmin) and BH mass at evaporation (Qmin) and the QCD background. Model: NGL with $M_P$ = 1 TeV .
\label{SPt Xmin}}
\end{figure}

In the p-p collisions BHs are formed when a given minimum mass is reached. This can be varied in
CATFISH using the  parameter $X_{min} \geq 1$ by $$M_{min}(formation) = X_{min}\times M_P.$$ Subsequently BHs evaporate through the Hawking mechanism until a minimum mass is reached, $$M_{min}(evaporation) = Q_{min}\times M_P$$ with $Q_{min}\geq 1$. It is assumed in our simulations that the minimum BH mass at formation and at evaporation are the same, but in general these two values are independent.  Figure \ref{SPt Xmin} shows results for how the minimum BH mass at formation and at evaporation affect the hadronic signal.  The $P_T$ is plotted for a $M_P$ of 1 TeV and two values of $X_{min}=Q_{min}$ using the NGL model. It is observed that at low $P_T$ the BHs with lower initial mass have a higher signal. This is because the decay product of the BHs with larger masses would have a higher $P_T$ than that of BHs with lower masses. For values above 1.5 TeV/c the $P_T$ is about the same for both, showing that BH production is insensitive to this parameter for higher $P_T$.\\

So far it has been shown that the possibilities of detecting a BH using the sum of the transverse momentum depend strongly on the model of BH formation and evaporation. The NGL model provides optimistic chances of producing BHs, whereas the YN and YR models provide a more pessimistic expectation, because the BH events would not be seen above ordinary QCD background.
Note that this does not imply that BHs cannot be detected in particle trackers at mid-rapidity. In the next section we explore such a possibility by studying the detection of charged remnants, if they are formed.

\subsection{Signal from BH Remnants}

\begin{figure}[pb]
\centerline{\psfig{file=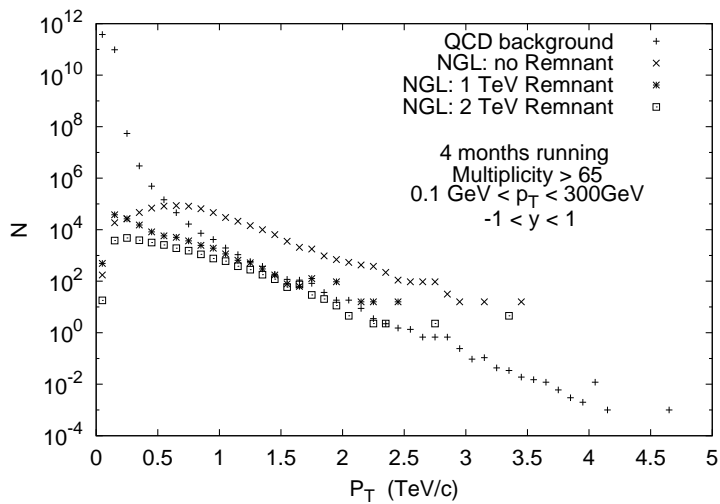,width=7.0cm,angle=270}}
\vspace*{8pt}
\caption{Comparison of the $P_T$ in the NGL model with no remnant and with BH remnants of 1 and 2 TeV
\label{BHRNGL}}
\end{figure}

The possibility of a BH not decaying completely to Standard Model particles after the Hawking evaporation phase and leaving a remnant has also been discussed in the literature \cite{BHremnant}. The BH remnant would be a very massive and possibly charged \cite{BHChargeRef} particle and these facts could be used to detect them in mid-rapidity particle trackers.
The remnants are produced in CATFISH when the evaporation phase is over and the mass has reached the value $M_{BHR} = Q_{min} \times M_P$, which is not further decayed into Standard Model particles.

\begin{figure}[pb]
\centerline{\psfig{file=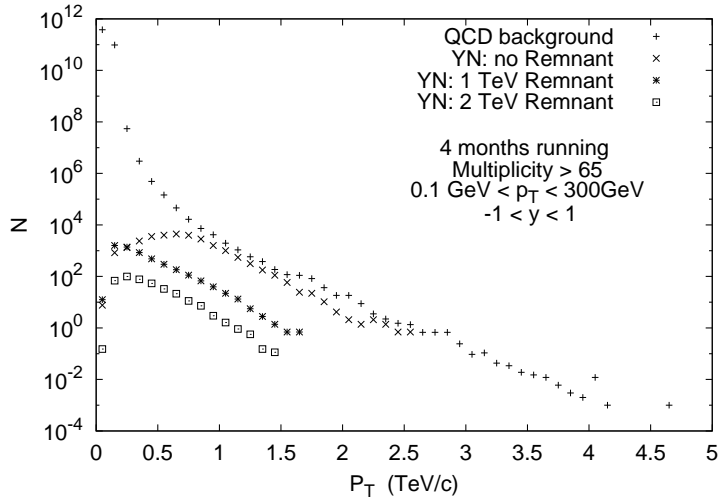,width=7.0cm,angle=270}}
\vspace*{8pt}
\caption{Comparison of the $P_T$ in the YN model with no remnant and with BH remnants of 1 and 2 TeV.
\label{BHRYN}}
\end{figure}

\begin{figure}[pb]
\centerline{\psfig{file=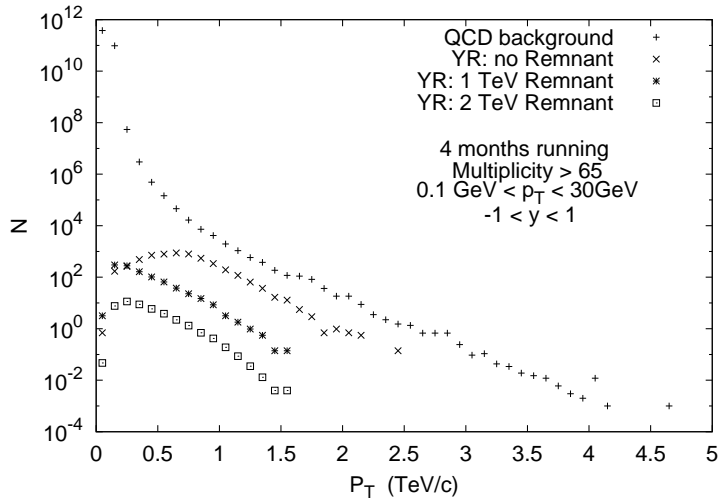,width=7.0cm,angle=270}}
\vspace*{8pt}
\caption{Comparison of the $P_T$ in the YR model with no remnant and with BH remnants of 1 and 2 TeV.
\label{BHRYR}}
\end{figure}

Using the technique of summing the transverse momentum it is seen in Figures \ref{BHRNGL},\ref{BHRYN} and \ref{BHRYR} 
that having a BH remnant significatively reduces  the total number of counts per bin.  
Even in the most optimistic scenario, the NGL model, Figure \ref{BHRNGL} shows that 
the BH remnant events are below the QCD background.  Therefore the possibility of detecting the BHs via the $P_T$ method are very small because the QCD background dominates.

\begin{figure}[pb]
\centerline{\psfig{file=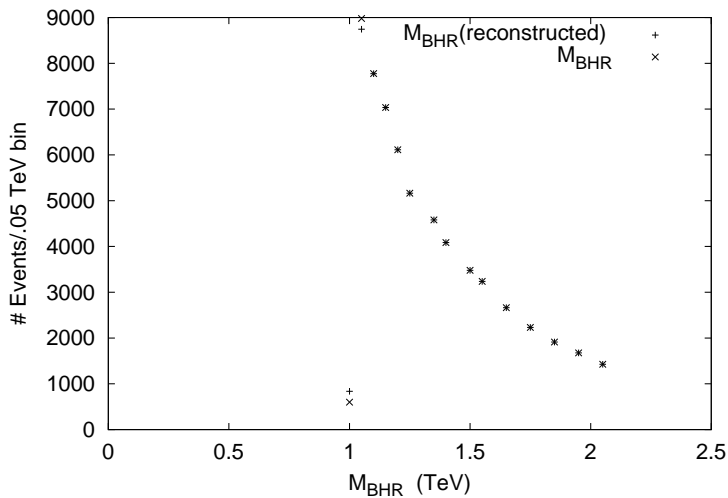,width=7.0cm,angle=270}}
\vspace*{8pt}
\caption{Comparison of BH remnant mass with the reconstructed mass from the TOF for $M_P$ = 1 TeV, $Q_{min}$ = 1
\label{BHRMass}}
\end{figure}

However, in light of having a BH remnant, another technique could be used to detect them. BH remnants may 
have a net charge \cite{BHChargeRef}, and that would allow one to detect these massive particles using the time-of-flight (TOF) method in particle trackers.  By knowing the momentum of the BH remnant and the TOF, its mass can be reconstructed using the following relation:

\begin{equation}
M_R=\frac{p'}{\gamma' \beta'},
\label{BHRMassEq} 
\end {equation}

\noindent where $p'$ is the momentum of the remnant. $\beta'$ and $\gamma'$ can be calculated respectively from

\begin{equation}
\beta'=\frac{x}{c t}, \gamma'=\frac{1}{\sqrt{1-\beta'^2}}
\label{Beta} 
\end {equation}

\noindent We assume a time resolution of 50 ps, a momentum resolution $\Delta p / p \approx 10\% p $ and a flight path of $x=$ 3 m.
In Figure \ref{BHRMass} we plot the counts per 0.05 TeV/c$ ^2$ bin of the actual BH remnant mass obtained from the simulation
and the reconstructed mass from the TOF and momentum. 
It can be seen that both graphs are almost the same for this bin size, thus demonstrating that the reconstructed mass is correct. 
The BH remnant mass is observed to have a distribution and not the exact value that was used in the simulation $Q_{min}\times M_P$.  As will be seen in Figure \ref{BHRComp}, the mass distribution becomes broader as the value of the minimum BH mass at evaporation gets closer to the Planck mass.

\begin{figure}[pb]
\centerline{\psfig{file=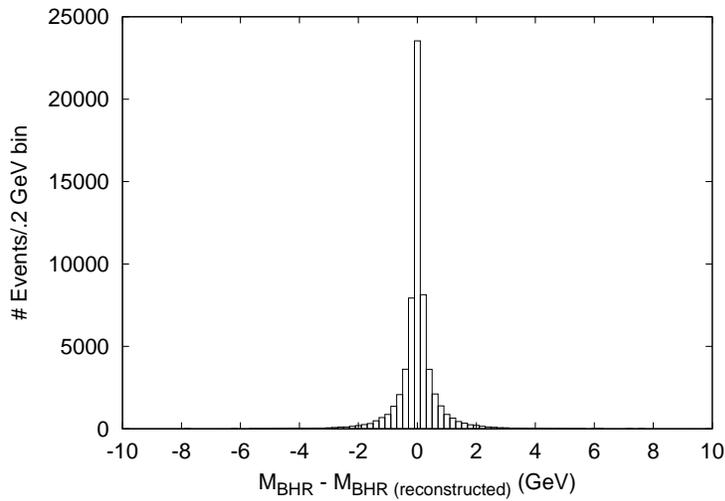,width=7.0cm,angle=270}}
\vspace*{8pt}
\caption{Difference in the BH remnant mass and the reconstructed mass for $M_P$ = 1 TeV, $Q_{min}$ = 1
\label{MassDif}}
\end{figure}
    
In Figure \ref{MassDif} we plot the difference between the actual mass of the BH and the reconstructed mass using a 0.2 GeV bin size.  The mass difference has an average of 0 GeV and a standard deviation of 0.76 GeV for this particular case, showing the high accuracy which is possible in determining the BH mass using a charged particle tracker plus TOF with reasonable performance characteristics.

\begin{figure}[pb]
\centerline{\psfig{file=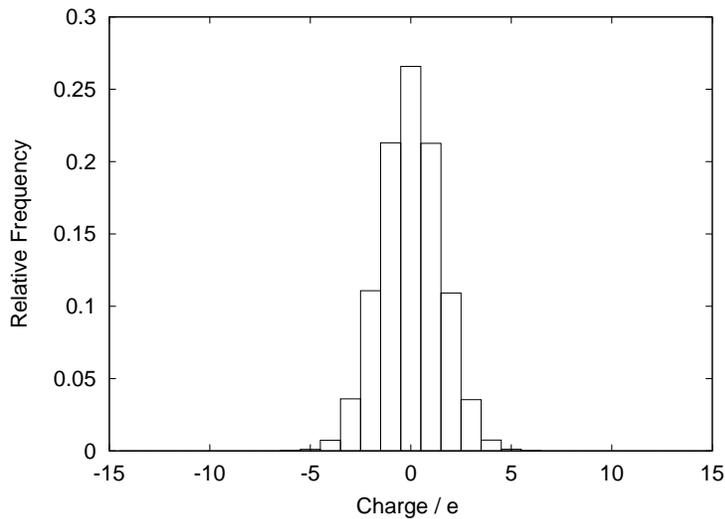,width=7.0cm,angle=270}}
\vspace*{8pt}
\caption{Charge Distribution among the BH remnants for $M_P$ = 1 TeV, $Q_{min}$ = 1
\label{BHCharge}}
\end{figure}

CATFISH does not assign a charge to the BH remnants and assumes they are all neutral, therefore in our simulations we assign a charge to each remnant, following reference \cite{BHChargeRef}. The charge distribution among the BH remnants is plotted in Figure \ref{BHCharge}. It is approximately Gaussian with  a standard deviation of 1.47.  The plot shows that about 75 percent of the BH remnants will have a net charge and could be detected using the tracking detector plus the TOF technique.

\begin{figure}[pb]
\centerline{\psfig{file=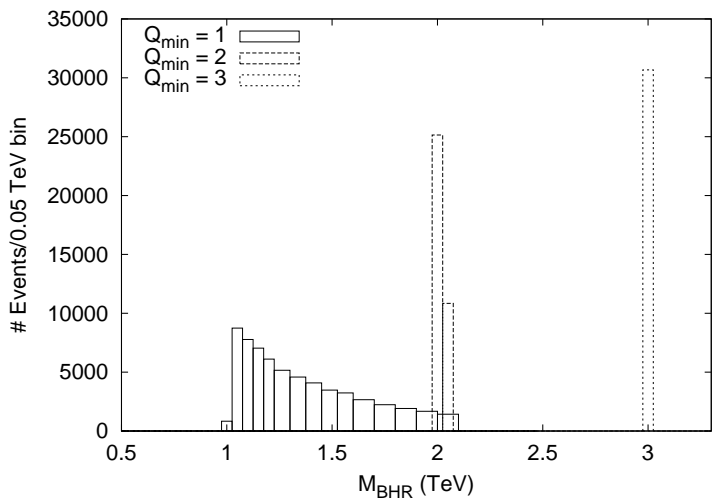,width=7.0cm,angle=270}}
\vspace*{8pt}
\caption{Comparison of different BH remnant mass distributions  for $M_P$ = 1 TeV.
\label{BHRComp}}
\end{figure}

As was mentioned earlier in the paper, the BH remnant mass has a distribution which is broader when it is closer to 
the Planck scale and it gets narrower when it is a few times larger. Figure \ref{BHRComp} shows a plot of how the BH 
remnant mass distribution changes depending on how high the minimum mass is above the Planck scale. One of the strong points of this method of detecting BHs is that the signal should be very clean in these large mass regions since no QCD process can generate such masses so that even small signals should be recognizable.

\section{Summary}

We have studied some of the possible methods which could be used to detect BHs at the LHC, if they are formed, by looking in the hadronic channel 
at mid-rapidity. Finding the sum of the total transverse momentum in
charged hadrons is one such method, because it is expected that the decay products of BH events have larger $P_T$ bin counts than normal QCD events at $P_T >$  0.5 TeV in the NGL model. The BH signature here is the transition from background QCD events to BH events. This model represents the upper bound for BH formation because there is no gravitational energy loss during formation. The other two models studied, the YN and YR models, take into account the energy loss at formation, and thus provide the lower bounds for the known models of BH formation. In this case lower counts per $P_T$ bin are expected and the hadronic signal will not be recognizable from the QCD background events. 
If the $P_T$ method does not work, another way of detecting BHs is possible if they do not decay entirely and, instead, leave a remnant. 
Using tracking detectors plus the TOF, the mass of a BH remnant can be reconstructed. By looking at the spectrum of masses from many events it could be easily recognized  that there is a very massive particle left over. This would be the signature for BH events with remnants happening at the LHC. 
There are other possible BH signatures which have not been considered here and have been studied elsewhere \cite{catfish,CF42,CF43,CF45,CF46,CF47,Cav}. Examples include the missing energy, missing transverse energy  and hadron energy signatures, as well as signatures due to suppression of back-to-back-correlated di-jets and di-lepton production with large transverse momentum .

\section*{Acknowledgments}

The authors wish to thank Professor Marco Cavagli\`a for his kind assistance in helping us run his code CATFISH. 
We also wish to thank Eric Anderson for corrections to the text and D.G Humanic for helpful suggestions. 
This work was supported by the U.S. National Science Foundation under grant PHY - 0653432.


\end{document}